\documentclass[12pt,manuscript]{aastex}
\usepackage{soul}

\slugcomment{}

\shorttitle{}
\shortauthors{}

\begin{document}

\title{X-RAY ENHANCEMENT AND LONG-TERM EVOLUTION OF SWIFT  J1822.3--1606}

\author{Onur Benli\altaffilmark{1}, 
        \c{S}. \c{C}al\i\c{s}kan\altaffilmark{1},
        \"{U}. Ertan\altaffilmark{1},
	M. A. Alpar \altaffilmark{1},
        J. E. Tr\"{u}mper\altaffilmark{2},
	N. D. Kylafis\altaffilmark{3}
}

\altaffiltext{1}{Sabanc\i\ University, Orhanl\i - Tuzla, \.Istanbul, 34956, Turkey}
\altaffiltext{2}{Max-Planck-Institut fuer extraterrestrische Physik, Giessenbachstrase, 85740 Garching bei Muenchen, Germany}
\altaffiltext{3}{Physics Department and Institute of Theoretical and Computational Physics, University of Crete, 71003 Heraklion, Crete, Greece}

\email{onurbenli@sabanciuniv.edu}

\def\be{\begin{equation}}
\def\ee{\end{equation}}
\def\ba{\begin{eqnarray}}
\def\ea{\end{eqnarray}}
\def\m{\mathrm}
\def\d{\partial}
\def\R{\right}
\def\L{\left}
\def\a{\alpha}
\def\acold{\alpha_\mathrm{cold}}
\def\ahot{\alpha_\mathrm{hot}}
\def\Mdotstar{\dot{M}_\ast}
\def\Omegastar{\Omega_\ast}
\def\OmegaK{\Omega_{\mathrm{K}}}
\def\Omegadot{\dot{\Omega}}
\def\Omegastardot{\dot{\Omega_\ast}}
\def\Mdotin{\dot{M}_{\mathrm{in}}}
\def\Mdotcrit{\dot{M}_{\mathrm{crit}}}
\def\Mdot{\dot{M}}
\def\nudot{\dot{\nu}}
\def\Edot{\dot{E}}
\def\Pdot{\dot{P}}
\def\Msun{M_{\odot}}
\def\Lin{L_{\mathrm{in}}}
\def\Lcool{L_{\mathrm{cool}}}
\def\Lacc{L_{\mathrm{acc}}}
\def\Rin{R_{\mathrm{in}}}
\def\rin{r_{\mathrm{in}}}
\def\rh{r_{\mathrm{h}}}
\def\rlc{r_{\mathrm{LC}}}
\def\rout{r_{\mathrm{out}}}
\def\rco{r_{\mathrm{co}}}
\def\Rout{R_{\mathrm{out}}}
\def\Ldisk{L_{\mathrm{disk}}}
\def\Lx{L_{\mathrm{x}}}
\def\Md{M_{\mathrm{d}}}
\def\cs{c_{\mathrm{s}}}
\def\dEb{\delta E_{\mathrm{burst}}}
\def\dEx{\delta E_{\mathrm{x}}}

\def\Bstar{B_\ast}
\def\Bb{\beta_{\mathrm{b}}}
\def\Be{\beta_{\mathrm{e}}}
\def\Rc{\R_{\mathrm{c}}}
\def\rA{r_{\mathrm{A}}}
\def\rp{r_{\mathrm{p}}}
\def\Bp{B_{\mathrm{p}}} 
\def\Tp{T_{\mathrm{p}}}
\def\Nh{N_{\mathrm{h}}}
\def\Md{M_{\mathrm{d}}}
\def\dMin{\delta M_{\mathrm{in}}}
\def\dM*{\delta M_*}
\def\Teff{T_{\mathrm{eff}}}
\def\Tirr{T_{\mathrm{irr}}}
\def\Firr{F_{\mathrm{irr}}}
\def\Tcrit{T_{\mathrm{crit}}}
\def\P0min{P_{0,{\mathrm{min}}}}
\def\Av{A_{\mathrm{V}}}
\def\ah{\alpha_{\mathrm{hot}}}
\def\ac{\alpha_{\mathrm{cold}}}
\def\tc{\tau_{\mathrm{c}}}
\def\p{\propto}
\def\m{\mathrm}
\def\fast{\omega_{\ast}}
\def\Alfven{Alfv$\acute{e}$n}
\def\418{SGR 0418+5729}
\def\142{AXP 0142+61}
\def\Caliskan{\c{C}al{\i}\c{s}kan}
\def\swift{Swift J1822.3--1606}
\def\sgr{SGR 0418+5729}
\def\psr{PSR J1734--3333}
\def\Gogus{G\"{o}\u{g}\"{u}\c{s}}


\begin{abstract}
We investigate the X-ray enhancement and the long-term evolution of the recently discovered, second \textquotedblleft low-B magnetar\textquotedblright ~\swift~ in the frame of the fallback disk model. During a soft gamma burst episode, the inner disk matter is pushed back to larger radii forming a density gradient at the inner disk. Subsequent relaxation of the inner disk could account for the observed X-ray enhancement light curve of  \swift. We obtain model fits to the X-ray data with basic disk parameters similar to those employed to explain the X-ray outburst light curves of other anomalous X-ray pulsars and soft gamma repeaters. The long period (8.4 s) of the neutron star can be reached by the effect of the disk torques in the long-term accretion phase ($1 - 3 \times 10^5$ yr). The currently ongoing X-ray enhancement could be due to a transient accretion epoch or the source could still be in the accretion phase in quiescence. Considering these different possibilities, we determine the model curves that could represent the long-term rotational and the X-ray luminosity evolution of \swift, which constrain the strength of the magnetic dipole field to the range of $1-2 \times 10^{12}$ G on the surface of the neutron star.

\end{abstract}


\section{ INTRODUCTION}  

The soft gamma repeater (SGR)  \swift~ was recently discovered (Cummings et al. 2011, \Gogus~ et al. 2011) as the second \textquotedblleft low-B magnetar\textquotedblright ~with $B \sim 2 \times 10^{13}$ G inferred from the dipole torque formula (Livingstone et al. 2011, Rea et al. 2012). Modeling the timing noise effects, Scholz et al. (2012) estimated that $B \sim 5 \times 10^{13}$ G with the same torque assumption. The first such source, the  \sgr,  indicates a magnetic dipole field of  $6\times 10^{12}$ G on the surface (equator) of the neutron star assuming that the source is spinning down by the dipole torques (Rea et al. 2013). This can actually be taken as an upper limit to the strength of the dipole field. If the neutron star is evolving with an active fallback disk, the dipole field strength that can produce the properties of this source could be in the $1 - 2 \times 10^{12}$ G  range on the surface of the neutron star (Alpar et al. 2011). These results clearly show that soft gamma bursts of anomalous X-ray pulsars (AXPs) and SGRs do not require magnetar ( $B > 10^{14}$ G) dipole fields. The properties of these two SGRs, which are likely to be older than the other known AXP/SGRs (see Mereghetti 2008 for a recent review of AXP/SGRs), provide tight constraints for models in explaining the long-term luminosity and the rotational evolution of AXP/SGRs in accordance with their statistical properties, like luminosity, period and period derivative distribution at different ages, and with possible evolutionary connections to the other young neutron star populations. 

For most AXPs and SGRs, the X-ray luminosity, $ \Lx$, is much higher than the rotational power, $\Edot = I \Omega \Omegadot$ where $I$ is the moment of inertia, and $\Omega$ and $\Omegadot$ are the angular frequency and the angular frequency derivative of the neutron star, respectively. In the magnetar model (Thompson \& Duncan 1995), the source of X-ray luminosity is the magnetic field decay, while in the fallback disk model (Chatterjee et al. 2000, Alpar 2001) the X-ray luminosity is produced by accretion onto the neutron star from the fallback disk and by intrinsic cooling when accretion is not possible.  Observed high-energy spectra of  AXP/SGRs can be explained by bulk-motion Comptonization in the accretion column of these sources (Truemper et al. 2010; 2013).  For a neutron star evolving with an active disk and a conventional dipole field, the dominant torque mechanism is the disk torque acting on the magnetic dipole field of the star.

At present, there are about 20 sources identified as AXP and SGR\footnote{%
http://www.physics.mcgill.ca/$\sim$pulsar/magnetar/main.html.}. All these sources have periods between 2 and 12 s, while the characteristic ages, $\tau = P/ 2 \Pdot$, vary from a few 100 to more than $10^7$ yr. Relatively long $P$ and large $\Pdot$ values of AXP/SGRs in comparison with those of normal radio pulsars place them to the upper right region of the $P - \Pdot$ diagram. Most of AXP/SGRs have not been detected in the radio band. Three sources that show pulsed radio emission have quite different radio properties from those of normal radio pulsars (Mereghetti 2013 and references therein). On the other hand, there are some radio pulsars in the AXP/SGR region of the $P - \Pdot$ diagram that might indicate an evolutionary connection between these \textquotedblleft high-B radio pulsars\textquotedblright and AXP/SGRs (see, e.g., Olausen et al. 2010). Among these radio pulsars, the recently measured braking index of \psr~ ($n = 0.9\pm 0.2$) shows that the source is evolving into the AXP/SGR region on the $P - \Pdot$ diagram (Espinoza et al. 2011).

In the magnetar model, the X-ray luminosity and the rotational properties of \sgr~ and \swift~ require rapid decay of the dipole component of the magnetic field. Models for field decay require a strong crustal toroidal field decaying together with the dipole field (Turolla et al. 2011, Rea et al. 2013). For \sgr, the initial toroidal field should be extremely strong ($4 \times 10^{16}$ G), while the initial dipole field should be $\sim 2-3 \times 10^{14}$ G to produce the current properties of the source (Turolla et al. 2011). For \swift, the model sources have initial toroidal and dipole fields of $4 \times 10^{14}$ G and $1.5 \times 10^{14}$ G, respectively (Rea et al. 2012). On the other hand, \psr~ seems to follow a completely different evolutionary path with an increasing period derivative. In the frame of the same model, this source could be in a short-term field growth phase. How this is related to the radio pulsar property and low X-ray luminosity of the source at a young age of  $\sim 10^4$ yr remains unclear. 

The long-term evolution model that we use in the present work was employed earlier to explain the properties of \sgr~  (Alpar et al. 2011) and  \psr~  (\Caliskan~ et al. 2013). According to these results, both \sgr~ and  \psr~ are evolving with fallback disks and conventional dipole fields ($B \sim 1- 5 \times 10^{12}$ G). The explanation of these apparently rather different sources, with the same model and with similar basic disk parameters, motivates us to further test the model by constraints provided by the sources with extreme properties. Here, we investigate the long-term evolution of the second \textquotedblleft low-B magnetar,\textquotedblright \swift. We try to determine the evolutionary epoch of the source and constrain the strength of the dipole field that gives consistent solutions for the long-term evolution.  We also try to explain the X-ray enhancement of \swift~ and discuss its possible effects on the current rotational properties of the source. We summarize the basic evolutionary stages of a neutron star evolving with a fallback disk and investigate the evolution of \swift~ in Section 2. A summary of our X-ray enhancement model and its application to the X-ray outburst light curve of \swift~ are given in Section 3. We discuss the results and summarize our conclusions in Section 4.

\section{ LONG-TERM EVOLUTION OF SWIFT J1822.3-1606}

For comparison with the model, we use the rotational properties and X-ray luminosity of \swift~ obtained from the most recent observational analysis of the source performed by Scholz et al. (2012). For an estimated distance of $\sim 1.6$ kpc, the quiescent bolometric luminosity of \swift~ is estimated between $2.5 \times 10^{31} - 2.6 \times 10^{33}$ erg s$^{-1}$. The period $P \sim 8.4 $ s (\Gogus~ et al. 2011). X-ray timing analysis, taking the noise effects into account, cannot constrain $\Pdot$ and gives $\Pdot \gtrsim 3 \times 10^{-13}$ s s$^{-1}$. Timing solutions with greater $\Pdot$ and marginally lower $\chi^2$ values could be obtained adding higher order derivatives of the frequency in the solutions to eliminate the noise effects (Scholz et al. 2012). In the accretion phase of the source, we adopt  $\Pdot \sim 4 \times 10^{-13} - 4 \times 10^{-12}$ s s$^{-1}$.  Details and applications of our long-term evolution model are given in Ertan et al. (2009), Alpar et al. (2011), and \Caliskan~et al. (2013). Here, we briefly describe the basic long-term evolutionary phases of a neutron star evolving with a fallback disk.

We follow the viscous evolution of an extended thin disk with an initial surface density profile $\Sigma = \Sigma_0 ~(\rin / r)^{3/4}$, where $\rin$ is the inner radius of the disk. Interaction of the inner disk with the magnetic dipole field governs the rotational evolution ($P, \Pdot, \ddot{P}$) of the neutron star. In the disk diffusion equation, we use the $\alpha$-prescription of the kinematic viscosity (Shakura \& Sunyaev 1973). Results of our earlier work on the enhancement light curves of transient and persistent AXP/SGRs imply that fallback disks of all these sources make a transition between hot and cold viscosity states at a critical temperature $\Tcrit \sim 1500 - 2000$ K. This model with $\a$ parameters $\ah \simeq 0.1$ and $\ac \simeq 0.045$ for the hot and cold viscosity states gives reasonable fits to observed X-ray enhancement light curves of different AXP/SGRs (\Caliskan~ \& Ertan 2012). We use the same $\a$ parameters in the long-term evolution of the disk. Note that (1) both $\ah$ and $\ac$ represent turbulent viscosities of an active disk, (2) the hot inner disk does not affect the long-term evolution of the disk, that is, the long-term history of the mass inflow rate of the disk is determined by the outer disk, and (3) the disk becomes passive at a temperature $\Tp \sim 100 - 200$ K (Ertan et al. 2009), which is much lower than $\Tcrit$. The transition temperature $\Tcrit$ between the viscosity states is important only for short-term events like X-ray enhancements (see Section 3), and should not be confused with $\Tp$ below, at which turbulent activity stops.  

All our simulations start with an outer disk radius $\rout = 5 \times 10 ^{14}$ cm. Subsequent evolution of the outer radius of the active disk is governed by the X-ray irradiation flux $\Firr = {C \Mdot c^2}/{4 \pi r^2}$ (Shakura \& Sunyaev 1973), where $c$ is the speed of light, $\Mdot$~ is the mass accretion rate onto the surface of the star, and the parameter $C$ represents the efficiency of X-ray irradiation. The X-ray luminosity is related to $\Mdot$~ through $L = {G M \Mdot}/{R}$ where $G$ is the gravitational constant and $R$ and $M$ are the radius and mass of the neutron star respectively. The mass-flow rate at the inner disk $\Mdotin = \Mdot / f $, where $f$ represents the fraction of $\Mdotin$ that is accreted onto the surface of the star. We take $f =1$; this simplification does not significantly affect our quantitative results. The analysis of the X-ray and infrared data of AXP/SGRs indicates that for all sources the irradiation efficiency $C$ is in the $1 - 7 \times 10^{-4}$ range for an inclination angle $i = 0\,^{\circ}$ between the normal of the disk and the line of sight of the observer  (Ertan \& \Caliskan~ 2006). 

Starting from the outermost disk, the disk gradually becomes passive as the local disk temperatures decrease below $\Tp$. The general properties of AXP/SGRs can be explained with $T_{p} \sim$ 100 - 200 K (Ertan et al. 2009). This is consistent with the results of the analysis indicating that the disk should be active even at 300 K (Inutsuka \& Sano 2005). In our long-term evolution model $\Tp$ and $C$ are degenerate parameters. In the model, if $C$ is increased from $1 \times 10^{-4}$ to $7 \times 10^{-4}$, a similar model curve can be obtained by increasing $\Tp$ only by a factor of $\sim 1.6$. That is, the lower and upper bounds on the range of $C$ obtained from earlier results (Ertan \& \Caliskan ~ 2006) remove the degeneracy and constrain $\Tp$ to values less than $\sim$200 K.

In recent years, observations and the results of theoretical calculations imply that magnetized neutron stars could accrete matter from the disk even in the fast-rotator phase (see, e.g., Rappaport et al. 2004). The critical value of the fastness parameter $\fast = \Omegastar / \OmegaK(\rA)$, above which accretion is completely hindered, is not well known. Here, $\OmegaK(\rA)$ is the angular velocity of the disk at the \Alfven~ radius, $\rA = (G M)^{-1/7}~\mu^{4/7}$ $\Mdotin^{-2/7}$, $ \Omegastar$ is the rotational angular frequency of the neutron star, and $\mu$ is the magnetic dipole moment of the neutron star. In the fallback disk model, AXP/SGRs are sources accreting in the spin-down phase. For these sources, $\rA$ is greater than the co-rotation radius, $\rco = (GM / \Omegastar^2)^{1/3}$. We employ the torque model obtained by Ertan \& Erkut (2008) through analysis of contemporaneous X-ray luminosity and period evolutions of XTE J1810--197 in the X-ray enhancement phase of the source. This spin-down torque acting on the star can be written as
\be
N = I ~\Omegastardot = \Mdotin ~(G M \rin)^{1/2} ~(1 - \fast^2),
\ee
where $\rin$ is the inner radius of the disk and $M$ is the mass of the neutron star which we take to be $1.4 \Msun$. In the accretion phase, this torque is equivalent to the integration of the magnetic torques from $\rA$ to $\rco$ taking the ratio of the poloidal and azimuthal components of the magnetic dipole field to be constant. When $\rA < \rlc = c/\Omegastar$, the light cylinder radius, we assume that the source is in the accretion phase and take $\rin = \rA$. It is implicitly assumed that the boundary layer of the disk could extend down to the co-rotation radius, while $\rin$ represents the inner radius of the thin disk or the outer radius of the boundary layer. For the accretion phase, from Equation (1) it is found that the period derivative, $\Pdot$, of the neutron star is independent of both $\Mdot$ and $P$ if the source is not close to rotational equilibrium, that is, when $\fast^2$ is not close to unity. The neutron star reaches maximum $\Pdot$ in the accretion phase. The maximum value of $\Pdot$ is proportional to $B_{0}^2$, where $B_0$ is the dipole field strength on the pole of the star. Depending on $B_0$, the initial period, $P_0$, and the disk mass, $\Md$, some of the model sources begin their evolution in the accretion phase, some of them enter the accretion phase at a later epoch, or others possibly never accrete over their entire lifetimes. The sources that cannot accrete are likely to be active radio pulsars as long as they remain above the pulsar deathline with their dipole field and current period. 

A fallback disk evolves to lower mass and $\Mdotin$.  In the long-term accretion phase, $\rin$ increases while the outer radius of the active disk $\rout$ decreases with gradually decreasing accretion rate, which brings about a decrease in the irradiation flux illuminating the disk. As more regions of the outer disk become passive, the outer radius of the active disk, $\rout$ moves inward. Propagation of $\rout$ inward also decreases the mass-flow rate from the outer to the inner disk regions. While $\rout$ approaches $\rin$, with rapidly decreasing $\Mdotin$, accretion luminosity enters the cut-off phase. The accretion luminosity first decreases below the cooling luminosity of the neutron star, and later accretion stops when the inner disk radius recedes to the light cylinder, $\rA = \rlc$. For the subsequent evolution, we take $\rin = \rlc$. In this phase, from Equation (1), it is found that $\Pdot \propto B^2 \Mdotin$. Even after the accretion phase, the disk torque prevails and continues to dominate the dipole torque, in most cases, over the visible lifetime of the sources. In this late phase of evolution, the disk torque and $\Pdot$ decrease with decreasing $\Mdotin$ converging to the level of the dipole torque, while $P$ remains almost constant. After the accretion has stopped, the X-ray luminosity is produced by the intrinsic cooling of the neutron star. In the total luminosity calculation, we include the cooling luminosity calculated by Page (2009) and the contribution of intrinsic dissipative heating due to the external (disk and dipole) torques acting on the star (Alpar 2007).

The important parameters of our model are $B_0$, $\Md$, $P_0$, and $\Tp$. Consistently with our earlier results, we keep $\Tp < 200$ K. Tracing $B_0$, $\Md$ and $T_p$, we obtain allowed ranges of these parameters that can produce the properties of \swift. Monte Carlo simulation of the radio pulsar population indicates that the initial periods could be represented by a Gaussian distribution centered around 300 ms with a width $\sim 150$ ms (Faucher-Giguere \& Kaspi 2006). In our simulations, we take $P_0 = 300$ ms. When we obtain a reasonable solution, we repeat the calculations with lower $P_0$ values to find the minimum $P_0$ value that allows the model source to enter the accretion phase and acquire the observed properties. 
        
Illustrative model curves that can represent the long-term evolution of Swift J1822.3-1606 are given in Figures 1 and 2. We investigate the evolutionary tracks considering the possibilities in quiescence: (1) the source is still in the accretion phase, and (2) the accretion phase terminated at an earlier time of evolution. For both cases, disturbances of the inner disk by soft gamma bursts could start a transient enhanced accretion epoch that can last for as long as decades. Considering this possibility, in Section 3, we also investigate whether the currently observed X-ray enhancement of the source could be produced by enhanced mass-flow rate of the disk caused by the soft gamma burst epoch that was observed just before the onset of the X-ray outburst (Rea et al. 2012). We discuss the possibilities for the quiescent-state properties of \swift~ considering the results of both the long-term evolution and the X-ray enhancement models in Section 4.

\section{X-RAY ENHANCEMENT OF SWIFT J1822.3-1606}

Starting in 2011 July, \swift~underwent an X-ray outburst (Rea et al. 2012).  The X-ray flux of the source in 1-10 keV range reached its maximum at t $\sim$ 0.76 days after the burst epoch. The subsequent smooth decay phase of the light curve was closely monitored by \textit{Swift, RXTE, Chandra, Suzaku} and \textit{XMM-Newton} for $\sim$ 400 days (Rea et al. 2012, Scholz et al. 2012).

The model we use in the present work is the same as that applied to the other AXPs and SGRs showing X-ray enhancements (see \c{C}al\i\c{s}kan \& Ertan 2012 for a detailed description of the model). The simple idea in this model can be summarized as follows: The fallback disk around the star has a thin disk profile in the quiescent state. A soft gamma-ray burst will push back part of the inner disk matter, which piles up at a larger radius, forming a density gradient at the innermost region of the disk. After the burst, starting with this new initial condition, relaxation of the disk to the pre-burst conditions with an enhanced mass-flow and accretion rate determines the X-ray luminosity evolution of the neutron star.

The pile-up at the inner disk is described by a Gaussian mass distribution, $ \Sigma = \Sigma_{\mathrm{max}}$ exp[$-(r-r_0)^{2}/\Delta r^{2}] $, centered at a radius $r_0$. We represent the extended thin disk by a power-law surface density profile, $\Sigma = \Sigma_{\mathrm{0}}~(r_{\mathrm{in}}/r)^{3/4}$. For the viscosity parameters ($\ah$, $\ac$), the irradiation efficiency ($C$), and the critical temperature ($\Tcrit$) that determines the transition between the viscosity states, we use similar values to those obtained for other AXP and SGRs (\Caliskan~ \& Ertan 2012).

For \swift,   $\rlc = 4 \times 10^{10}$ cm, and the results of the long-term evolution model imply that $\rin$ is close to $\rlc$ in quiescence.  We take the position $r_0$ of the pile-up to be outside the light cylinder, $r_0 > \rlc$. X-ray enhancement light curves of the model sources are not sensitive to the exact positions of $r_0$, $\rin$, and the details of the Gaussian distribution, but sensitive to initial relative positions of these radii and the total mass included in the Gaussian distribution (Ertan et al. 2006, \Caliskan~ \& Ertan 2012). In the simulations, we set $r_0 = 4.1 \times 10^{10}$ cm, leaving   $\Sigma_{\mathrm{max}}$ and $\Delta r$, which define the pile-up mass under the Gaussian, and $\rin$ as free parameters. The enhancement light curve is given in  Figure 3.

The abrupt decrease in the decay phases  of the model light curves seen in Figure 3 occurs when the radial position of the hot - cold viscosity border of the disk with radius $\rh$ approaches the inner disk radius. The innermost disk, which is rapidly depleted by the hot state viscosities, cannot be refilled by the matter in the cold viscosity state 
at the same rate, causing a sharp decrease in the luminosity. Subsequent refilling of the evacuated innermost disk leads to a small increase in the luminosity, followed by a smooth decay to the level of the cooling luminosity. This effect does not modify the X-ray light curve in the early decay phase, when $\rh$ is not very close to $\rin$, since the surface density gradients and local variations in the mass-flow rate occurring at larger radii are smoothed out on the way to the inner disk radius. A detailed investigation of this feature on the model light curves for different quiescent luminosities can be found in \Caliskan~\& Ertan (2012).

For comparison with data, we adopt a distance of 1.6 kpc for \swift~ (Scholz et al. 2012). In the luminosity calculation, we also add the contribution of the cooling luminosity. With $\Lcool$ = 4 $\times$ 10$^{32}$ erg s$^{-1}$, we obtain a better fit to the last data points (Figure 3). This is consistent with the estimated range of the quiescent luminosity of the source. For the $\a$ parameters of the viscosity, we use the same values ($\ah$~= 0.1 and $\ac$~= 0.045) employed in the long-term evolution model (Section 2) and in the X-ray enhancement models of other AXP/SGRs (\Caliskan~ \& Ertan 2012). In Figure 3, it is seen that the solid model curve is in good agreement with the X-ray data of \swift. For this model, $C$ = 3 $\times$ 10$^{-4}$ and $\Tcrit$ = 1500 K. The cooling luminosity (horizontal line) and the accretion luminosity, $\Lacc$, curves are also presented separately in Figure 3. The accretion continues even after $\Lacc$ decreases below $\Lcool$ and converges to its quiescent level, or it could stop at a critical $\Lacc$ depending on the dipole field strength and the current period of the source. For the allowed $B_0$ range obtained from the long-term evolution model, and with the condition $\rA = \rlc$ for termination of accretion, this critical accretion luminosity could be $\sim 10^{30}$ erg s$^{-1}$. 

\section{DISCUSSION AND CONCLUSIONS}

Comparing the results of the X-ray enhancement and the long-term evolution models, we try to understand the properties of \swift~ in the quiescent state. In our long-term evolution model, the magnitude and $\Mdotin$ dependence of the disk torque are quite different during and after the accretion phase. In the accretion phase, the disk torque given in Equation (1) has a weak dependence on $\Mdotin$, and $\Pdot$ remains constant close to its maximum value provided that the source does not come close to rotational equilibrium. For the model sources given in Figures 1 and 2, the time intervals with constant $\Pdot$ correspond to this phase. In the model, the accretion phase terminates when $\rA = \rlc$, afterward $\rin$ remains attached to $\rlc$. For this phase, from Equation (1), we obtain $\Pdot \p \Mdotin$. 

Our results show that the X-ray enhancement light curve of \swift~ could be explained by additional enhanced mass flow from the inner disk onto the surface of the star. This does not guarantee that the source is in the accretion phase in the quiescent phase. It is possible that the source could currently be in a transient accretion phase that could come to an end when the accretion luminosity decreases below a critical level depending on the dipole field strength of the source. To sum up, from the observed properties of the source in the current enhancement phase, we do not know whether the star accretes matter in quiescence.

At present, in the enhancement phase, the evolution of $\Pdot$ depends on the state of the source in quiescence: (1) if the star is  accreting matter in the quiescent state, the currently observed $\Pdot$ remains roughly constant while the accretion luminosity is gradually decreasing to the quiescent level, (2) if the source does not accrete in quiescence, it is now in a transient accretion phase; accretion stops at a critical $\Mdotin$, and subsequently $\Pdot$ starts to decrease with decreasing $\Mdotin$. Note that, for case (2),  $\Pdot$ of the source in quiescence could be much lower than in the enhancement phase. 

First, we consider the case that the source was not in the accretion phase before the onset of the X-ray enhancement.   In Figure 1, the illustrative model curves with $B_0$ values $1.74$ and $3.0\times 10^{12}$ G give very high $\Pdot$ values in the accretion phase and could be eliminated. Let us consider the long-term evolution represented by the dotted (red) curve with $B_0 = 1 \times 10^{12}$ G in Figure 1. The long-term accretion phase of this model source ends at point  B shown on the $\Pdot$ curve (bottom panel).  Note that the period is less than the observed period (8.4 s) of \swift~ at point B. In the quiescent phase, the source could be at a point on the decay phase of $\Pdot$, say at point A seen on the $\Pdot$ curve with $P \simeq 8.4$ s. After the burst epoch, with enhanced mass-flow rate, the inner disk penetrates the light cylinder, which begins accretion, thus causing the onset of the X-ray enhancement phase.  $\Pdot$ increases to a level close to its maximum value reached earlier during the long-term accretion phase.  As long as the inner disk remains inside the light cylinder ($\rin = \rA$),   $\Pdot$ remains constant. After $\Mdotin$ decreases below the level corresponding to the critical condition $\rA = \rlc$,  accretion stops, while $\Pdot$ decreases gradually with decreasing $\Mdotin$ converging to its quiescent level indicated by point A on this illustrative model curve (Figure 1).  

Here, we assume that in the transient accretion phase, the disk can build up the boundary layer conditions prevailing in the steady-state long-term accretion phase. A transient accretion may take place with a less efficient disk torque acting on the star  than in the long-term accretion epoch. In this case, observed rotational properties do not give information about the maximum $\Pdot$ of the long-term evolution, and none of the model curves given in Figure 1 can be excluded, since they all reproduce the period and the X-ray luminosity of  \swift~ simultaneously.     

Alternatively, in quiescence,  the source could already be in the accretion phase. If that in the case, then we expect that the source will continue to remain in the accretion state until the long-term accretion phase terminates without a significant change in $\Pdot$. The model curve given in Figure 2 illustrates a long-term evolution of this type. This model source continues to accrete matter when $P$ reaches the observed value. Now,  it is close to the end of the long-term accretion phase with an accretion luminosity that remained  below the cooling luminosity a few $10^4$ yr ago. For this alternative scenario, our model cannot produce the source properties if the actual $\Pdot$ is less than $\sim 6 \times 10^{-13}$ s s$^{-1}$. 

The condition $\rA = \rlc$ to terminate the accretion phase is a simplification of our model. The accretion could stop earlier when $\rA$ increases gradually to a certain ratio of $\rlc$ and the phase with $\Pdot \p \Mdotin$ could start at an earlier epoch. This does not affect the qualitative features of the long-term model curves, but increases the critical mass-flow rate of the disk below which we expect significant torque variations with decreasing $\Mdotin$. In the late decay phase of the enhancement, the accretion luminosity,  $\Lacc$, decreases and the viscous timescale of the inner disk increases. It is not possible to observe  the decrease in $\Lacc$ once it falls below $\Lcool$. The expected decrease in $\Pdot$ would be observed without a significant accompanying change in the X-ray luminosity. If $\rA = \rlc$ is a close representation of the condition for the onset of accretion, it may take decades for $\rA$ to reach $\rlc$ and terminate accretion in view of the long period of the source (8.4 s), which gives $\rlc = c P /2\pi \simeq 4 \times 10^{10}$ cm. If accretion stops for a smaller $\rA$, the effect could be observable within years.

In alternative models (see, e.g., Pons \& Rea 2012), the X-ray enhancement light curve of \swift~ is produced by cooling of the neutron star's crust that is heated by the mechanism producing soft gamma bursts. 

We have shown that both the X-ray enhancement light curve and the long-term evolution of \swift~ can be explained in the frame of the fallback disk model. Main disk parameters used in the present work are very similar to those employed in our earlier work to explain the X-ray enhancements and the evolution of AXP and SGRs (Ertan et al. 2009, Alpar et al. 2001, \Caliskan~ \& Ertan 2012, \Caliskan~ et al. 2013). The model sources with the dipole field strength in the $ 0.5 - 1.0 \times 10^{12}$ G range on the pole of the star and with the initial periods greater than $\sim 55$ ms can reach the X-ray luminosity and the rotational properties of \swift~ simultaneously.

We conclude that the source is accreting matter from the disk at present in the X-ray enhancement phase. Accretion could persist after the enhancement phase with the accretion luminosity decaying to quiescent level that could remain below the cooling luminosity. A more interesting possibility is that the mass transfer onto the star could stop at a critical accretion rate, and subsequently the disk torque and the $\Pdot$ of the source could start to decrease as proportional to decreasing mass-flow rate arriving at the inner disk. In this phase, since the accretion luminosity remains below the cooling luminosity, variation in $\Pdot$ could be observed without a significant change in the observed luminosity of the source. 

\acknowledgements

We acknowledge research support from Sabanc\i\ University and from
T\"{U}B{\.I}TAK (The Scientific and Technical Research Council of
Turkey) through grant 110T243. M.A.A. is a member of the Science Academy, Istanbul, Turkey. We thank Ersin G\"{o}\u{g}\"{u}\c{s} for useful discussions.

\clearpage

\clearpage 
\begin{figure}[t]
\centerline{\includegraphics[width=0.9\textwidth,angle=270]{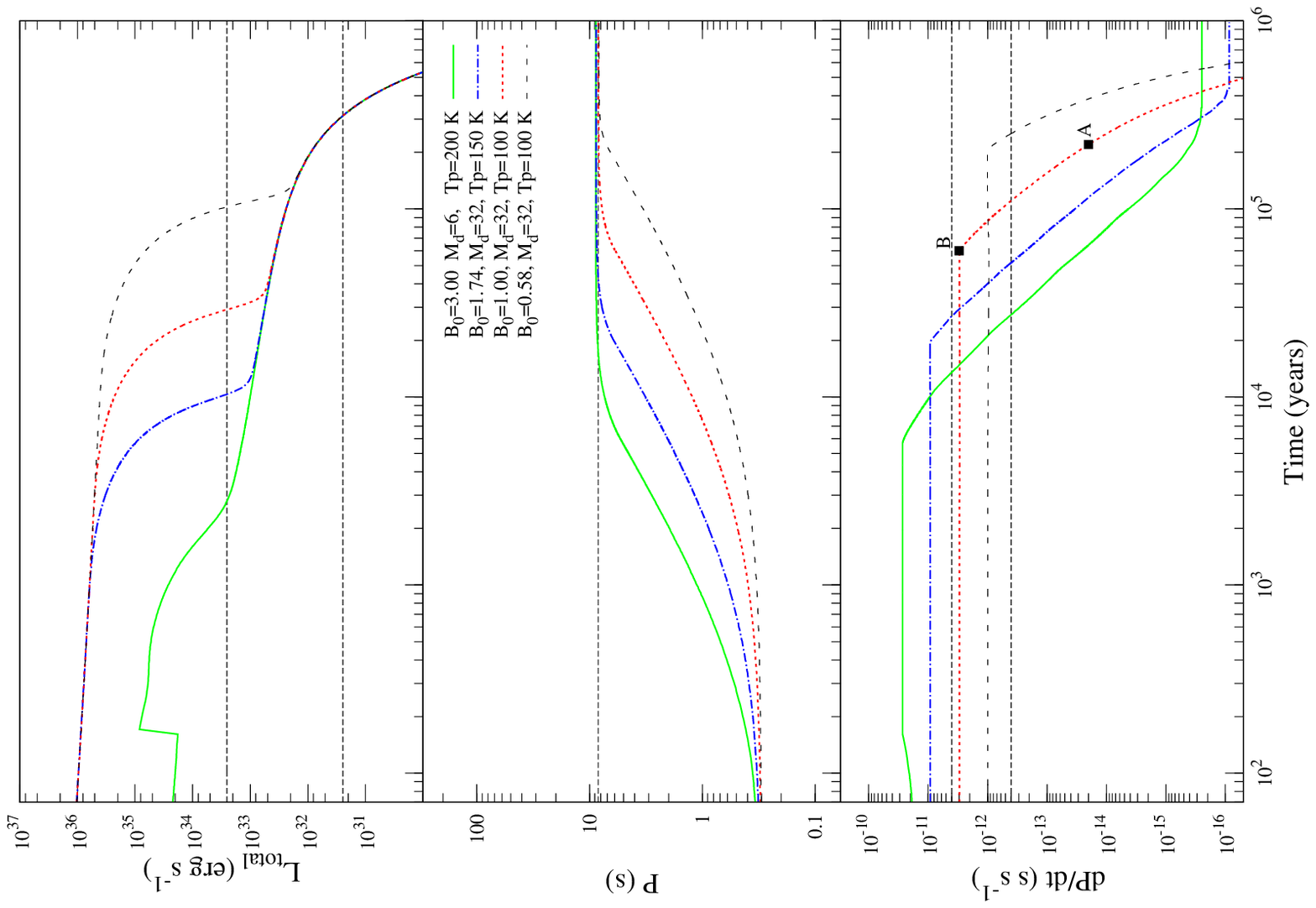}} 
\caption{\label{fig:1}
Long-term luminosity, period and period derivative evolution of the model sources.  The values of $\Md$  and $\Bp$ employed in the models  are given in the middle panel in units 10$^{-6}$ $M_{\sun}$ and 10$^{12}$ G respectively. Horizontal lines in the top panel show the observational error bars of $L$ (Scholz et al. 2012). For these models, $C$ = 1 $\times$ 10$^{-4}$ except for the dashed (black) line which has $C$ = 7 $\times$ 10$^{-4}$ and  represents the evolution with minimum $B_0$. In the bottom panel, horizontal lines show the range of $\Pdot$ for \swift~ adopted in the present work. The model sources with $B_0$ values 0.58 and 1.00 $\times 10^{12}$ G could represent the evolution of  \swift~ if the source is not in the accretion phase in quiescence. The points A and B seen in the bottom panel are for a discussion of possibilities for the source properties in quiescence  (see the text for explanation). }
\end{figure}

\clearpage 
\begin{figure}[t]
\centerline{\includegraphics[width=1\textwidth,angle=270]{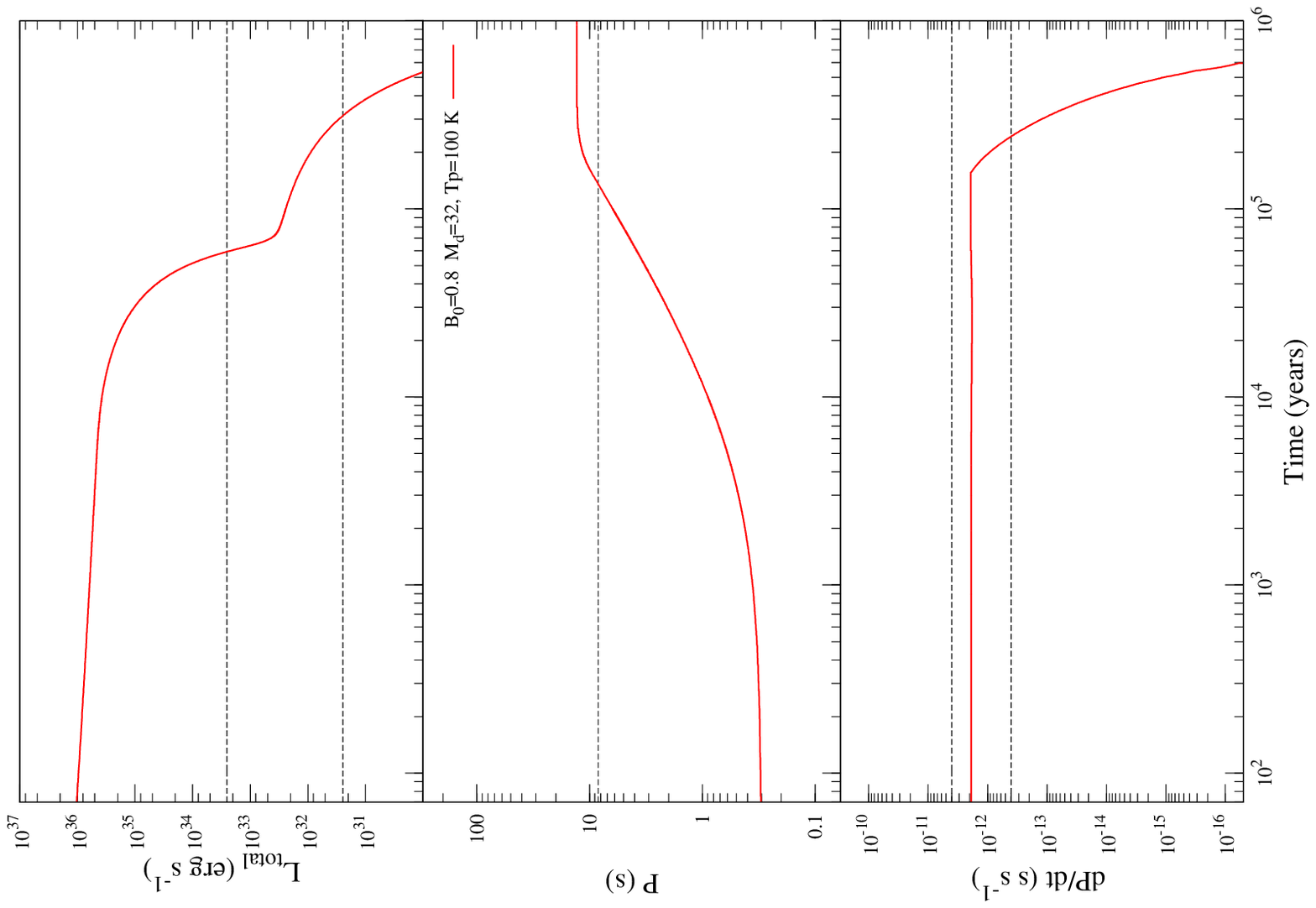}} 
\caption{\label{fig:2}
Evolution of an illustrative model source which could acquire the properties of \swift~ simultaneously if the source  is in the long-term accretion phase at present in the quiescent state. The values of $\Md$  and $B_0$ are given in the figure in units of  10$^{-6}$ $M_{\sun}$ and 10$^{12}$ G respectively. For these models we take $C = 3 \times 10^{-4}$.}  
\end{figure}

\clearpage

\begin{figure}[t]
\centerline{\includegraphics[width=0.75\textwidth,angle=270]{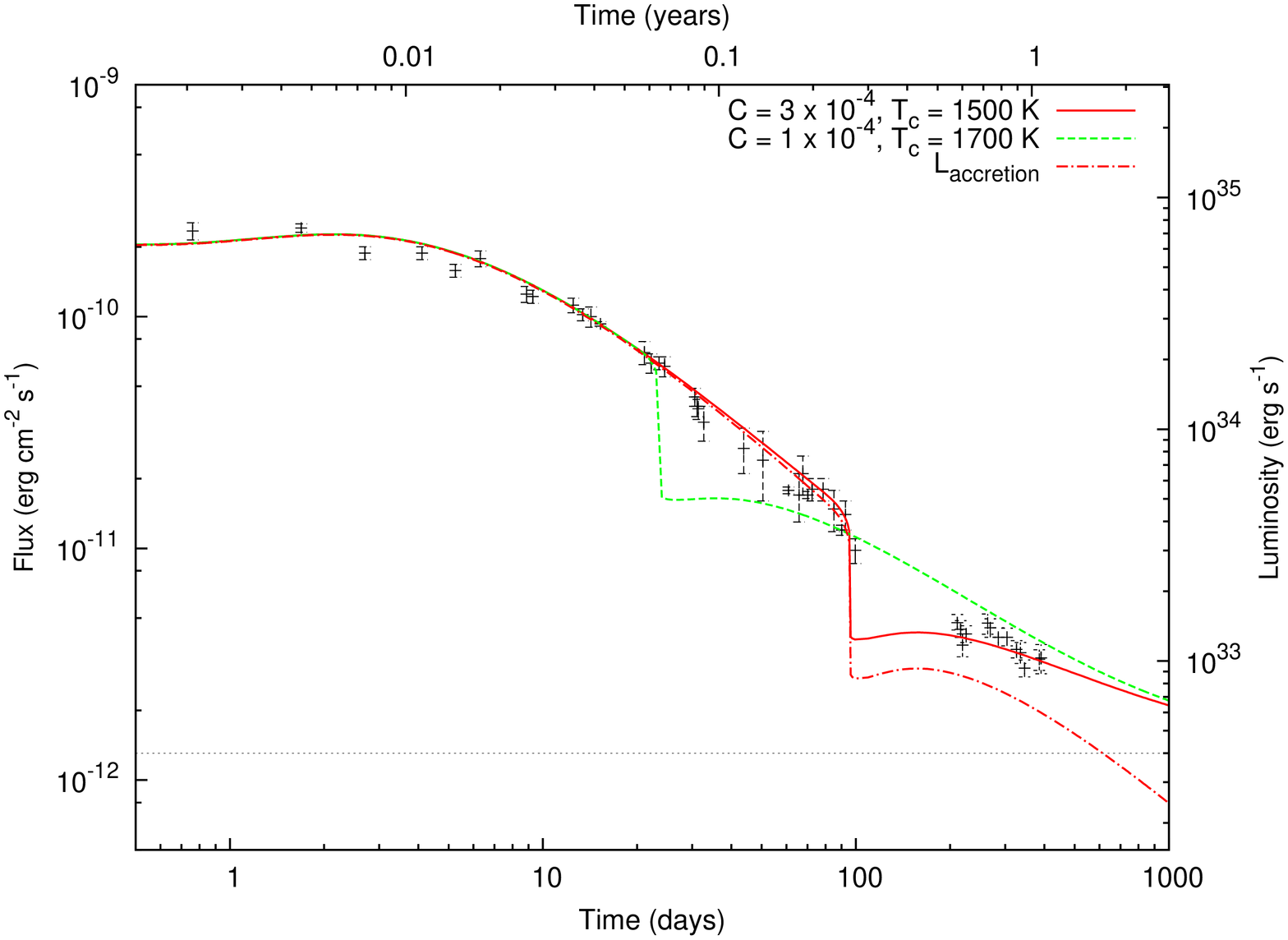}}
\caption{\label{fig:psds} 
X-ray enhancement data of \swift. The data is taken from Rea et al. (2012) and Scholz et al. (2012). The luminosity is given on the right axis, assuming a distance of 1.6 kpc. Values of the parameters $\Tcrit$ and $C$ are given in the figure. For both models $\ah = 0.1$ , $\ac = 0.045$ and $B_0 = 1 \times 10^{12}$ G. The abrupt decrease seen in the model curves are produced when the innermost disk enters the cold viscosity state (see the text for explanation). The accretion luminosity (dot-dashed line) and the level of cooling luminosity taken in the models (horizontal line)  are also given separately in the figure.}
\end{figure}

\end{document}